\documentclass[journal=jacsat,manuscript=article]{achemso}

\usepackage[version=3]{mhchem} % Formula subscripts using \ce{}
\usepackage{nameref}
\usepackage{color}
\usepackage{siunitx}
\usepackage{comment}
\author{Henri Truong}
\affiliation{Univ. Bordeaux, CNRS, Centre de Recherche Paul-Pascal (CRPP), UMR 5031, 115 Avenue Schweitzer, 
F-33600 Pessac, France}
\author{Chiara  Moretti}
\affiliation{CNRS, ENSL, Laboratoire de Chimie, UMR 5182, 46 all\'ee d’Italie, F-69364 Lyon, France}

\author{Lionel Buisson}
\affiliation{Univ. Bordeaux, CNRS, Centre de Recherche Paul-Pascal (CRPP), UMR 5031, 115 Avenue Schweitzer, 
F-33600 Pessac, France}

\author{Benjamin Ab\'ecassis} 
\affiliation{CNRS, ENSL, Laboratoire de Chimie, UMR 5182, 46 all\'ee d’Italie, F-69364 Lyon, France}

\author{Eric Grelet} 
\email{eric.grelet@crpp.cnrs.fr}
\affiliation{Univ. Bordeaux, CNRS, Centre de Recherche Paul-Pascal (CRPP), UMR 5031, 115 Avenue Schweitzer, 
F-33600 Pessac, France}

\title{Light-Activated Self-thermophoretic Janus Nanopropellers
}


\begin{document}

%%%%%%%%%%%%%%%%%%%%%%%%%%%%%%%%%%%%%%%%%%%%%%%%%%%%%%%%%%%%%%%%%%%%%
%% The abstract environment will automatically gobble the contents
%% if an abstract is not used by the target journal.
%%%%%%%%%%%%%%%%%%%%%%%%%%%%%%%%%%%%%%%%%%%%%%%%%%%%%%%%%%%%%%%%%%%%%
\begin{abstract}

Achieving controlled and directed motion of artificial nanoscale systems in three-dimensional fluid environments remains a %central
key-challenge in %the field of 
active matter, primarily due to the prevailing thermal fluctuations
that rapidly randomize the particle trajectories. 
%In this low Peclet number regime (Pe $\sim$~1), where self-propulsion and Brownian diffusion are of comparable %of the same order of 
%magnitude, demonstrating and controlling particle activity %proves particularly challenging/
%is particularly difficult.  
While significant progress has been made with micrometer-sized particles, imparting sufficient mechanical energy, or self-propulsion, to nanometer-sized particles to overcome Brownian diffusion %thermal noise effects
and enable controlled transport remains a major %obstacle 
issue for emerging applications in nanoscience and nanomedicine. 
Here, we address this challenge by demonstrating the fuel-free, reversible, and tunable active behavior of %70~nm 
gold-silica (Au-SiO$_2$) Janus nanoparticles (radius $R\sim$~33~nm) induced by optical excitation. Using single particle tracking, we provide direct experimental evidence of self-thermophoresis, clearly distinguishing active motion from thermal noise. %Our results reveal that light-driven self-propulsion accounts for around 60\% of the enhanced dynamics of the Janus nanoparticles.  
These light-driven Janus nanoparticles constitute a minimal yet robust photothermal system for investigating active matter and its manipulation at the nanoscale.

\begin{comment}
**************************
Abstract of 150 words:\\
Achieving controlled motion of artificial nanoscale systems in 3D fluid environments is a major challenge in active matter due to significant thermal fluctuations. While progress has been made with micrometer-sized particles, imparting sufficient energy to nanometer-sized particles to overcome Brownian diffusion and enable controlled transport remains an issue for nanoscience and nanomedicine. Here, we demonstrate the fuel-free, reversible, and tunable active behavior of 66~nm gold-silica (Au-SiO2) Janus nanoparticles %(radius $R\sim$~33~nm) 
induced by optical excitation. Using single particle tracking, we provide direct experimental evidence of self-thermophoresis, clearly distinguishing active motion from thermal noise. These light-driven Janus nanoparticles offer a minimal yet robust photothermal system for investigating and manipulating active matter at the nanoscale.
\end{comment}
\end{abstract}
\clearpage
%%%%%%%%%%%%%%%%%%%%%%%%%%%%%%%%%%%%%%%%%%%%%%%%%%%%%%%%%%%%%%%%%%%%%
%% Start the main part of the manuscript here.
%%%%%%%%%%%%%%%%%%%%%%%%%%%%%%%%%%%%%%%%%%%%%%%%%%%%%%%%%%%%%%%%%%%%%
\section{Introduction}

%Brownian active matter refers to microscopic units, that convert energy from their environment into mechanical energy for self-propulsion. This term encompasses a wide variety of both living and artificial systems, such as suspensions of bacteria \cite{schwarz-linek_escherichia_2016, wensink_meso-scale_2012}, cytoskeletal filaments \cite{julicher_active_2007} and protein motors \cite{kodera_video_2010,sanchez_spontaneous_2012}, Janus self-phoretic colloids \cite{paxton_motility_2005} ... \\
%In recent years, the interest for simple artificial active systems with tunable self-propulsion at the colloidal and nanoscale has grown. This field has garnered considerable attention due to its potential applications, including drug delivery, precision nanosurgery, biopsy and cell sorting  \cite{wang_nanomicroscale_2012, patra_intelligent_2013, nelson_microrobots_2010, gao_environmental_2014, ebbens_active_2016, abdelmohsen_micro-_2014, maier_optical_2018}. However, the design of active particles that can be effectively controlled at the nanoscale remains a challenge. \\
%\red{GENERAL REFS to ADD}    
Active Brownian matter, which converts external energy into autonomous motion at micro- and nanometric dimensions, provides unique opportunities to investigate nonequilibrium dynamics and develop functional devices at small scales. Natural examples such as motile bacteria and cytoskeletal filaments powered by motor proteins, illustrate how energy transduction %at small scales 
can generate complex directed behavior \cite{schwarz-linek_escherichia_2016, wensink_meso-scale_2012,julicher_active_2007, kodera_video_2010,sanchez_spontaneous_2012}. Inspired by these biological systems, recent advances in nanotechnology have enabled the development of synthetic micro- and nanopropellers that harness chemical fuels, light, or electromagnetic fields to propel themselves through fluids \cite{ghosh_controlled_2009,Buttinoni2012,Vutukuri2020,He2024}.
While the controlled propulsion of micrometer-sized particles is now well established \cite{jiang_active_2010}, achieving reliable and tunable motion at the nanoscale remains a major challenge. At these dimensions, Brownian motion rapidly randomizes trajectories, hindering sustained directional transport in three-dimensional environments \cite{lee_self-propelling_2014}. To design and characterize  effective nanoscale propellers, it is therefore crucial to understand how random thermal fluctuations compete with and limit directed motion at these small scales \cite{Buhler2025}. Micro- and nanoparticles undergo passive Brownian motion described by their translational diffusion coefficient, D$_{\rm{T}}$ = $k_B T/6 \pi\eta R$, and rotational diffusion coefficient, D$_{\rm{R}}$ = $k_BT/8 \pi \eta R^3$, where $R$ is the particle radius, $k_\mathrm{B}$ the Boltzmann constant, $T$ the absolute temperature, and $\eta$ the solvent viscosity. As particle size decreases to the nanoscale, random thermal diffusion becomes stronger, counteracting any directed motion, while the propulsion efficiency typically decreases, as it is usually related to the particle surface area.  
To compare the effects of directed propulsion, with velocity $v$, to stochastic Brownian motion, the dimensionless P\'eclet number (Pe) is commonly used \cite{Bechinger2016}, %This quantity reflects, over the characteristic length \(L\) of the system, the relative importance of the active motion described by the particle velocity \(v\) against its diffusive passive behavior described by D$_{\rm{T}}$ i.e. \(Pe = \frac{L \cdot v}{\mathrm{D_T}}\) that can also 
defined as $Pe \propto \frac{ v}{\sqrt{\mathrm{D_T \cdot D_R}}}$. %with $v$ the particle velocity.
%This nanoscale also significantly complicates the quantitative characterization of particles dynamics by requiring even more sensitive measurement techniques.
%As particle size decreases, self-propulsion weakens, making it increasingly difficult to characterize active behavior beyond experimental error margins and provide a definitive proof of active behavior in nanoscale systems. 
When $Pe \sim 1$, active and passive contributions are comparable, making the clear demonstration and control of nanoscale propulsion particularly demanding.
To address this challenge, %to enhance self-propulsion at the nanoscale, 
%Better understanding of active matter at the nanoscale is needed, and various 
%Overcoming this constraint is essential for applications that require precise nanoscale manipulation, including targeted drug delivery, nanomedicine, environmental remediation, and sensing.
various nanopropeller designs and actuation strategies have been proposed, tailored to specific applications such as targeted drug delivery, precision nanosurgery, biopsy, and related fields of nanomedicine \cite{wang_nanomicroscale_2012, patra_intelligent_2013, nelson_microrobots_2010, gao_environmental_2014, ebbens_active_2016, abdelmohsen_micro-_2014, maier_optical_2018,Su2023,Ruiz-Gonzalez2025}. 
Among these approaches, light-driven propulsion is particularly promising, as it enables fuel-free, %non-toxic, 
reversible, and tunable control of nanoswimmer behavior \cite{Lin2017,Palagi2019,Sipova2020,Calero2020,Liu2021,Schmidt2021,Kollipara2023,Rey2023,Ciriza2023,Braun2024,He2025}. 
Building on this concept, we demonstrate %fuel-free, reversible and tunable
self-thermophoretic propulsion of gold–silica (Au-SiO$_2$) Janus nanoparticles %(made of two different faces)
under optical excitation. Our Janus particles, with a radius of $R\sim$~33~nm, consist of 
spherical gold cores partially coated with an asymmetric silica shell. Using single-particle tracking via optical microscopy, we provide direct experimental evidence that these visible-light-activated nanopropellers generate sufficient mechanical energy to overcome Brownian diffusion, with active and passive contributions to their dynamics being comparable ($Pe \sim 1$). To isolate these contributions, we analyze the motion of Janus nanoparticles relative to that of bare gold nanoparticles under identical conditions, enabling direct characterization and quantification of the tunable active self-propulsion, distinct from the passive component arising from hot Brownian motion.
%By comparing the dynamics of Janus nanoparticles to that of bare gold nanoparticles under identical optical conditions, we directly characterize and quantify the tunable active behavior, distinguishing it from the passive contribution due to hot Brownian motion. 
Our study establishes a minimal yet robust system for investigating and manipulating active matter at the nanoscale.

\section{Materials and methods}
\label{mat_methods}
\subsection{%Experimental systems
Synthesis and characterization of Au-SiO$_2$ Janus nanoparticles}

\begin{figure}
\centering
\includegraphics[width=\textwidth/2]{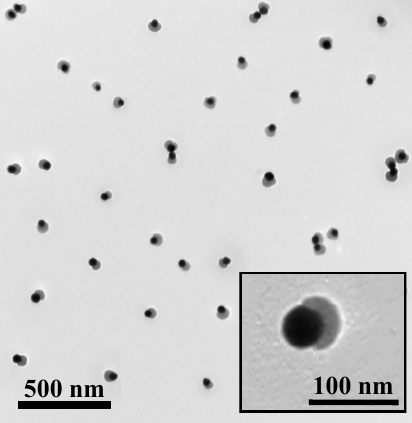}
\caption{\textbf{Transmission electron microscopy (TEM) image of Au-SiO$_2$ Janus nanoparticles}. The inset displays a high-magnification view of an individual Janus nanoparticle, revealing the distinct contrast between the silica shell (light gray) and the Au nanobead (black).}
\label{Janus_AuNPs}
\end{figure}

\begin{figure}
\centering
\includegraphics[width=1\textwidth]{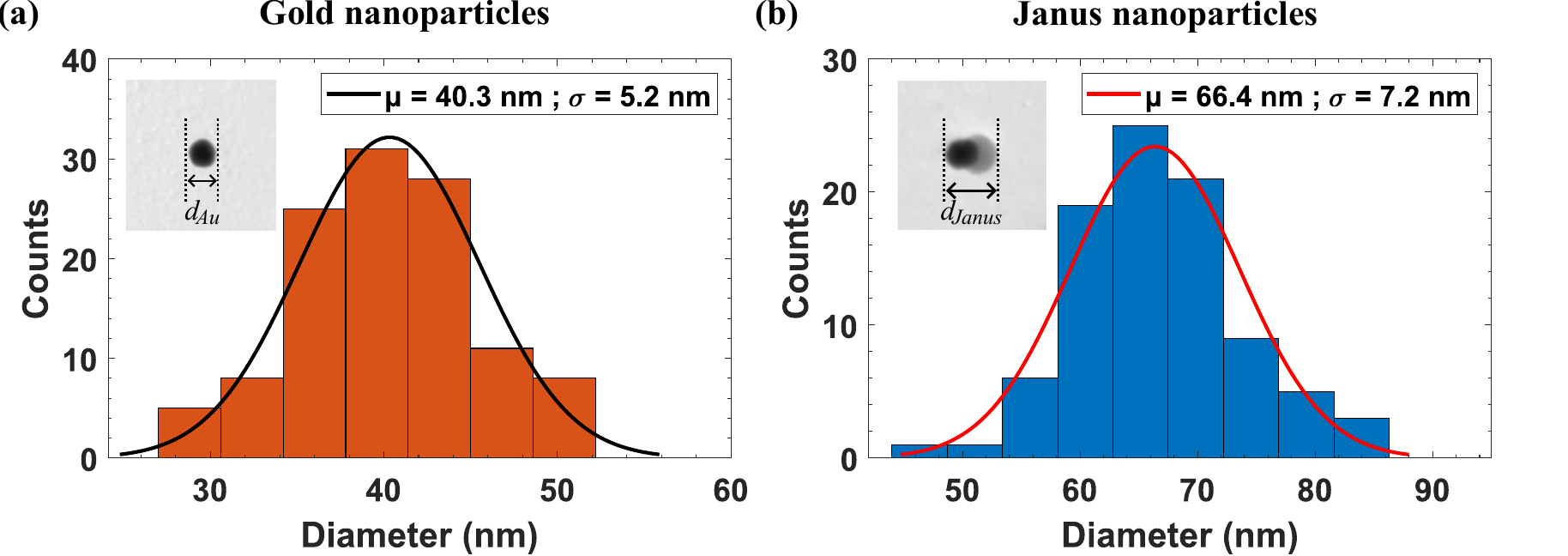}
\caption{\textbf{Nanoparticle size distribution determined by transmission electron microscopy (TEM)}. (a) Bare gold nanobeads (N = 116) and (b) Janus nanoparticles (N = 90). The gold nanobeads are nearly spherical, characterized by a diameter $d_{Au}$, while $d_{Janus}$ refers to the major axis of the Janus nanoparticles (see inset). Solid lines represent Gaussian fits to the measured size distributions;  error bars on the particle diameter correspond to the standard deviation of these fits: $d_{Au} = 40.3 \pm $ 5.2~nm and $d_{Janus} = 66.4 \pm 7.2$~nm.}
\label{size_distribution}
\end{figure}

Gold-silica Janus nanoparticles are synthesized via selective silica (SiO$_2$) nucleation on one hemisphere of gold nanospheres \cite{castro_solution_2016}. The Janus morphology is confirmed by transmission electron microscopy (TEM), as shown in Fig.~\ref{Janus_AuNPs}.
Specifically, the gold nanoparticles are synthesized using the multi-step protocol described by Bastus \textit{et al.} \cite{bastus_kinetically_2011}. This results in gold nanobeads with an average diameter of $d_{Au} = 40.3 \pm 5.2$~nm, as determined from the size distribution shown in Fig.~\ref{size_distribution}(a). The selective growth of a silica shell on one hemisphere of the Au nanoparticles is performed following the procedure published by Chen \textit{et al.} \cite{chen_scalable_2010} and by Castro \textit{et al.}\cite{castro_solution_2016}, using an alcohol/water solution as the reaction medium. Surface functionalization is achieved by grafting two competing ligands: polyacrylic acid (PAA) and 4-mercaptophenylacetic acid (MPAA). Due to their chemical incompatibility, these surfactants undergo phase separation, forming distinct patches of homogeneous composition on the nanoparticle surface. The subsequent addition of tetraethyl orthosilicate (TEOS) induces the selective nucleation of porous silica on the MPAA functionalized regions, resulting in the formation of Au-SiO$_2$ heterodimers. The silica shell has an average thickness of about 25~nm (Fig.~\ref{Janus_AuNPs}), producing Janus nanoparticles with a major axis of $d_{Janus} = 66.4 \pm$7.2~nm (Fig.~\ref{size_distribution}(b)), as determined by TEM. 
For TEM analysis, nanoparticles %with a concentration of %C$\_{\rm{TEM}} \sim 
($\sim 10^9$ particles/mL), are deposited onto O$_2$ plasma-treated carbon-coated grids. Imaging is performed using a Hitachi H-600 electron microscope operating at 75~kV, equipped with an AMT CCD camera for image acquisition.

\begin{figure}
\centering
\includegraphics[width=0.6\textwidth]{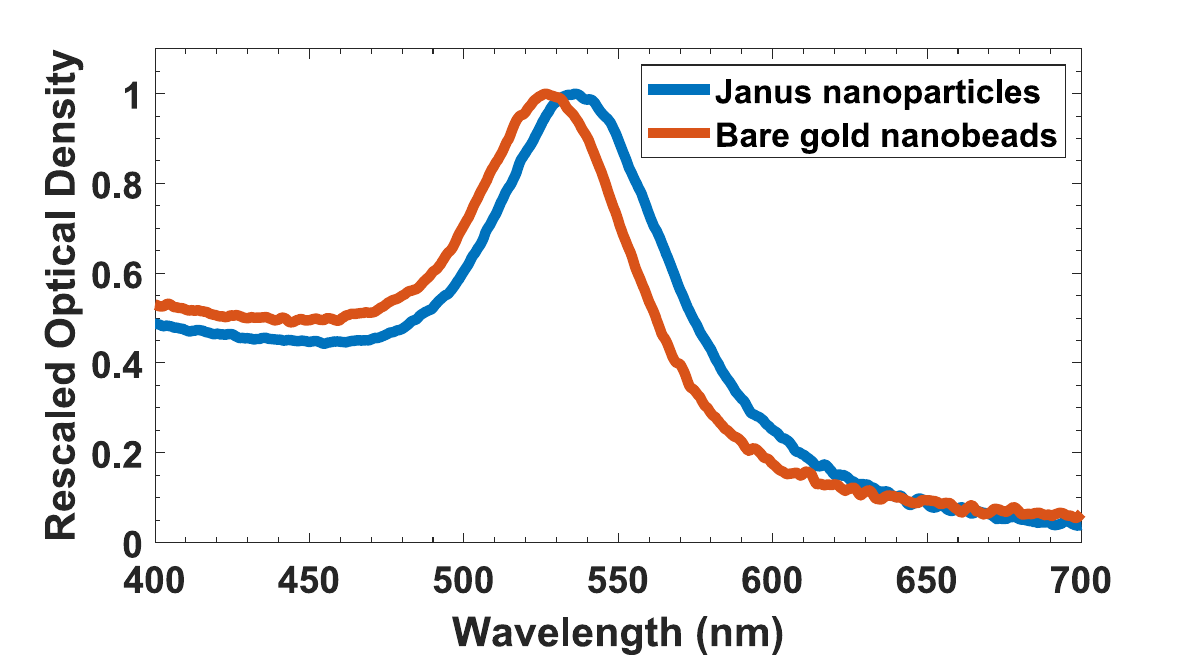}
\caption{\textbf{Visible absorption spectra} of Janus nanoparticles (blue line) and bare gold nanobeads (orange line). For comparison purposes, the optical density of each spectrum has been normalized to its respective maximum absorption value to account for the difference in sample concentration. The corresponding maximum optical densities are OD$_{max,Au}$ =~3.56 and OD$_{max,Janus}$ =~2.99. }
\label{OD_spectrum}
\end{figure}

The visible light absorption spectra of both Janus and Au nanoparticle suspensions are recorded using a Nanodrop One spectrophotometer (Thermo Scientific) (Fig.~\ref{OD_spectrum}), enabling concentration determination. The Janus nanoparticle suspension exhibits a red shift in the plasmon resonance peak from $\sim$530~nm to $\sim$540~nm. This shift is attributed not only to the growth of the silica shell, but also to changes in the local environment due to ligand exchange, notably the presence of thiol groups bound to the nanoparticle surface \cite{Amendola2017}. It is worth mentioning that nanoparticle suspensions exhibit excellent colloidal stability, with no observable aggregation over several months. 
Before thermopheresis experiments, suspensions are dialysed against a propionate buffer (pH 5.5, ionic strength I = 0.5 mM).

\subsection{Optical setup}

The sample is visualized using a Dark-Field (DF) microscopy setup equipped with an Olympus UPlanFLN 100$\times$ oil immersion objective with an adjustable numerical aperture (NA$_{\rm{objective}} =$ 0.6 -- 1.3) and an Olympus U-DCW oil immersion dark-field condenser (NA$_{\rm{condenser}} $ = 1.2~--~1.4). 1000 images per video are acquired using a Hamamatsu ORCA-Quest qCMOS camera C15550-20U at an effective pixel size of 2$\times$4.6~µm (Binning 2) and a frame rate of 476~fps.

\begin{figure}
\centering
\includegraphics[width=\textwidth]{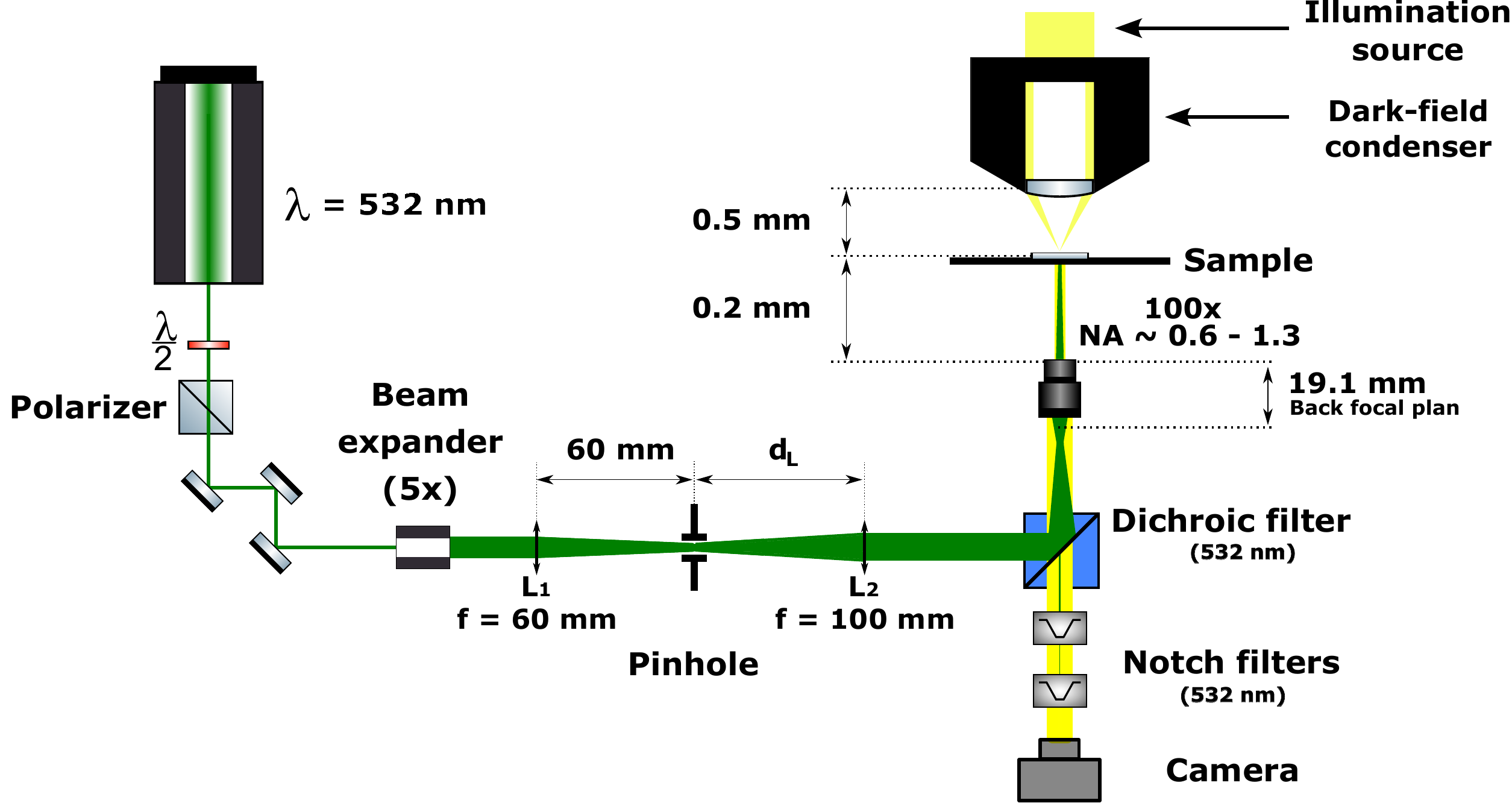}
\caption{\textbf{Schematic representation of the optical setup} enabling both direct visualization of the sample 
via dark-field microscopy and its excitation using green laser illumination. }
\label{setup}
\end{figure}

The optical excitation targeting an homogeneous illumination of the sample is carried out using a Spectra-physics Excelsior 532 Single Mode laser ($\lambda$ = 532~nm, maximum nominal output power: 200~mW). To protect the camera from laser exposure, two notch filters are placed in the imaging path. The incident laser power on the sample is measured and calibrated before each experimental session using a photodiode (Thorlabs S121C) positioned at the sample level. The laser power is modulated using a Glan–Thompson prism, acting as polarizer, paired with a rotating half-wave ($\lambda/2$) plate. The beam size is first increased using a Thorlabs GBE05-A 5$\times$ beam expander, and then further enlarged through a series of lenses (L1 and L2) to ensure full illumination of the objective's back focal plane (Fig.~\ref{setup}). To reduce light intensity gradient across the sample area, %the spatial concentration of the laser beam and minimize its effect on the sample, 
the final lens in the optical path is axially translated by a few millimeters (distance d$_{\rm{L}}$ in Fig.~\ref{setup}), generating controlled beam divergence before the back focal plane of the objective. This slight defocusing broadens the beam profile, thereby enhancing illumination uniformity, with an estimated beam size of about 130~µm. Under these conditions, no optical artifacts, including nanoparticle optical trapping, are detected, even at the maximum incident laser power of P~=~81~mW on the sample (after accounting for optical losses in the setup). 
%Bean size defined at Imax/e^2 corresponding to 2 sigma (standard deviation): r~65µm.
%Normalized by the beam size, this corresponds to a maximum laser power density of about $1000$~W/cm$^2$ at the sample level.
%This step is essential to ensure the most uniform laser irradiation of the sample and in first order no effect coming from it has been characterized. 
\\
%This configuration yields a beam diameter of about 200~µm at the sample plane. %corresponding to an illuminated area of S $\simeq 31.4 \times$10$^3$ µm$^2$.

\subsection{Sample preparation and Single Particle Tracking (SPT)}

\begin{figure}
	\includegraphics[width=\columnwidth]{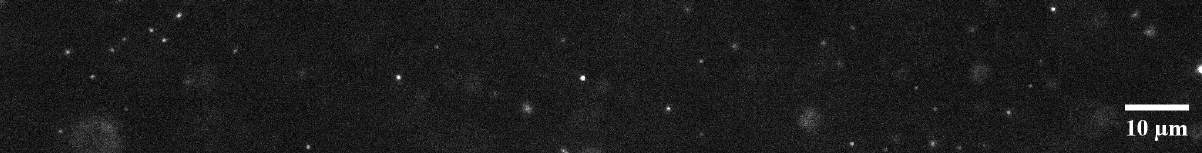}
	\caption{\textbf{Dark-field microscopy image of the Janus nanoparticles}. A representative frame acquired during the experiments (see Supplementary Movie S1 in the SI), showing Janus nanoparticles in the absence of optical excitation. Although individual nanoparticles are subject to optical resolution limits, this does not prevent the precise determination of their center-of-mass position via single particle tracking (see Material and methods). The sample concentration is set in the dilute regime at c~$\approx$ 3.6$\times$10$^9$ particles/mL.}
    \label{illus_DF_Janus}
\end{figure}

For thermophoresis experiments, the particle concentration is set in the dilute regime at c$\sim$ 3.6$\times$10$^9$ particles/mL (OD$_{\mathrm{520nm}}$ = 0.05), as shown in Fig.~\ref{illus_DF_Janus}. 
Microscope cells for imaging the samples are prepared by thoroughly cleaning glass slides and 0.17~mm-thick cover glasses (18 × 18 mm) with chromosulfuric acid to remove organic contaminants. Subsequently, both glass components are treated with an O$_2$ plasma cleaner (Basic Plasma Cleaner, Harrick Plasma) to enhance surface hydrophilicity. A 4~µL aliquot of the particle suspension is then deposited at the center of the glass slide, and the cover glass is gently placed on top of the droplet. The edges are immediately sealed with optical adhesive (Norland Optical Adhesive NOA 68), producing an observation chamber approximately 10~µm thick.
%The suspension is loaded between a glass slide and a coverslip, which are sealed using optical glue (Norland Optical Adhesive NOA 68), forming an observation chamber with a thickness of $\sim 10$~µm. 
The mean inter-particle distance is estimated as \(<l> \sim n^{-1/3}\), where \(n = N/V\) is the particle number density, yielding %\(<l> \sim 6 \, \mu m\)
$<l> \sim 6$~µm  
at this concentration. Given the nanoparticle diameter,  %\(d \simeq  60\, nm\), 
this corresponds to a mean inter-particle distance \(\frac{<l>}{d}\) of approximately 100 times the particle size. Under these conditions, no collective behavior is expected from a colloidal standpoint, and the measured diffusion coefficients are expected to match those at infinite dilution. 

\begin{figure}
	\includegraphics[width=0.97\columnwidth]{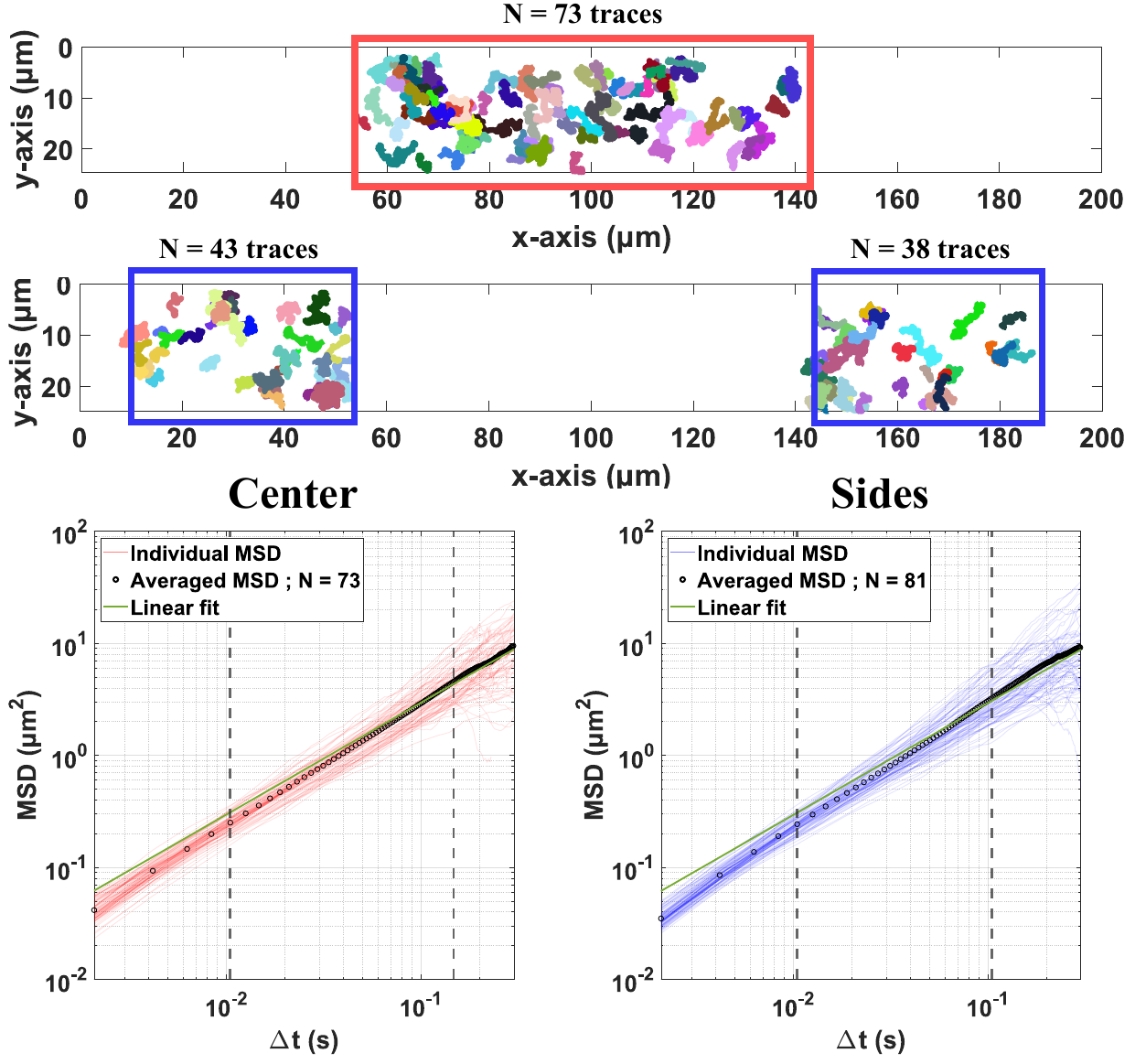}
	\caption{\textbf{Uniform illumination and Janus nanoparticle dynamics}.
    Trajectories of Janus nanoparticles tracked using dark-field microscopy (see Fig.~\ref{illus_DF_Janus}) under maximum laser illumination (P~=~81~mW) are shown for particles located inside (red) and outside (blue) the central Region of Interest (ROI) (top and middle panels). The areas of the red and blue boxes are equal, resulting in similar numbers of traces (N=73 for the central ROI and N=81 [43+38] for the sides), allowing therefore a direct comparison of the dynamics. The corresponding mean squared displacement (MSD) curves are displayed (bottom panel), with red and blue curves corresponding to particles inside and outside the central ROI, respectively. Linear fits to the mean MSD curves yield diffusion coefficients of D$_{\textrm{T,Center}}$ = 7.31 µm$^2$/s and  D$_{\textrm{T,Sides}}$ = 7.26 µm$^2$/s. The minimal difference between these values, which falls within the statistical error, confirms the spatial homogeneity of laser illumination across the field of view. Data represent averages over four independent acquisitions.
    }
    \label{homogeneity}
\end{figure}

Image acquisition is performed at the middle height of the observation cell, maintaining a distance of approximately 5~µm away from all surfaces to minimize surface-particle interactions. A initial reference experiment is performed without laser irradiation, followed by measurements under laser illumination, and concluded with a final control experiment. Prior to each acquisition at a given laser intensity, the sample is illuminated for 5~min  to reach thermal steady state. Two-dimensional particle trajectories $\textbf{r}(t)$ are determined using a custom-written MATLAB (MathWorks) particle-tracking algorithm, adapted from the method developed by Crocker and Grier 
\cite{crocker_methods_1996}. Although the nanoparticles are smaller than the optical resolution limit, their center-of-mass positions are localized with sub-diffraction precision by fitting their diffraction patterns with a 2D Gaussian function \cite{Shen2017}.
Particle dynamics is characterized by calculating their Mean Squared Displacements (MSD), at lag time $\tau$, defined as $\mathrm{MSD}(\tau) = \langle [\mathbf{r}(\tau+t)-\mathbf{r}(t)]^2 \rangle_t$ where $\left \langle ... \right \rangle_t$ denotes the average over all start times $t$. The MSD is calculated for each trace, before being first averaged over the total number of detected particles (around one hundred per experiment, see SI) and subsequently fitted to determine the corresponding effective diffusion coefficients (Fig.~\ref{homogeneity} and Figs.~S1 \& S2) \cite{Novotny2019,Mallouk2021}. %Due to depth of focus leading to the loss of the particles tracking outside the plane of focus, we assume the recorded dynamics to be mainly restricted to a 2D visualization.\\

\section{Results and discussion}

A batch of gold-silica Janus nanoparticles with a nominal diameter of approximately 66~nm has been produced via selective silica (SiO$_2$) nucleation on one hemisphere of gold nanospheres (see \nameref{mat_methods} for more details). Bare gold nanobeads serve as an experimental control system to isolate the active contribution of the Au-SiO$_2$ Janus nanoparticles under optical excitation. 

Particle dynamics is investigated in the dilute regime using Single Particle Tracking (SPT). Our custom setup combines Dark-Field (DF) microscopy \cite{lee_self-propelling_2014} -- which leverages the strong scattering signal of gold for high-contrast visualization -- with simultaneous optical excitation of the sample via a defocused green laser beam ($\lambda = 532$~nm) delivered through the same objective (Fig.~\ref{setup} and \nameref{mat_methods}). 
A representative DF image of the Janus nanoparticles is provided in Fig.~\ref{illus_DF_Janus}. Illumination homogeneity is rigorously controlled to eliminate artifacts induced from light gradients (Fig.~\ref{homogeneity}).

\begin{figure}
\centering
\includegraphics[width=\textwidth]{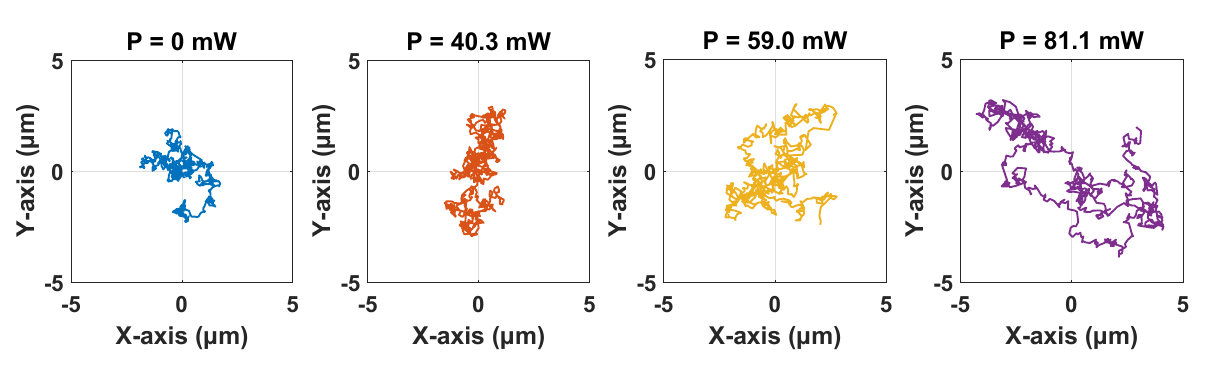}
\caption{\textbf{Representative trajectories of Janus nanoparticles under increasing optical excitation power}, illustrating the self-thermophoretic active contribution to their dynamics over a fixed trace duration of 1.8~s. %All the shown trajectories have same duration of 1.8~s. and are centered for illustration purpose.
}
\label{Janus_traces}
\end{figure}

\begin{figure}
\centering
\includegraphics[width=\textwidth]{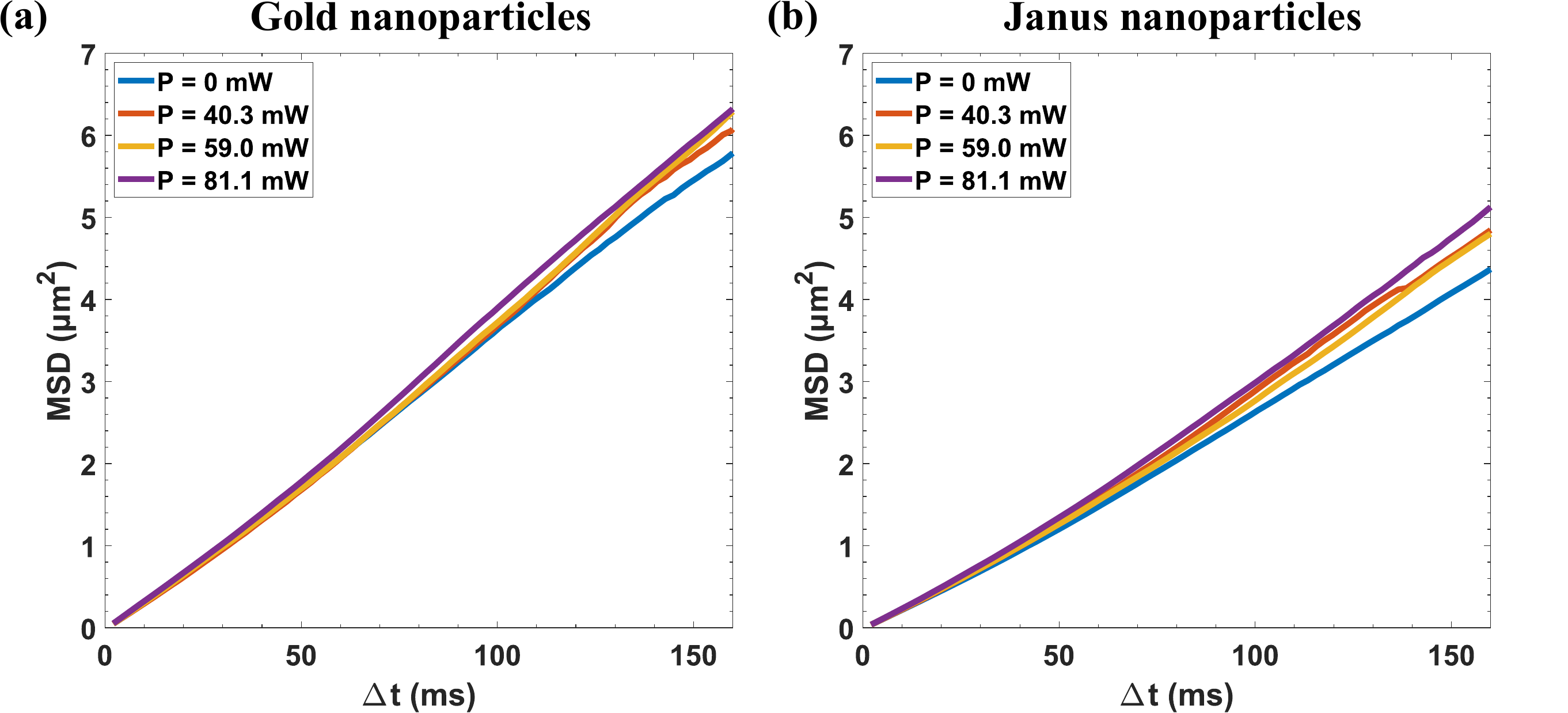}
\caption{\textbf{Ensemble-averaged mean squared displacements (MSDs)} for (a) bare gold nanospheres and (b) Janus nanoparticles under increasing optical power. Data represent averages over about 100 individual particles (see Figures S1 and S2). In both systems, the increasing slopes of the MSD curves with higher illumination intensity indicate enhanced effective diffusion coefficients.}
\label{MSD_linlin_AuNPs_Janus}
\end{figure}

Particle trajectories are captured as two-dimensional (2D) projections within the microscope focal plane. Due to the finite depth of field of the objective, these 2D trajectories result from the projection of the particle's full three-dimensional (3D) motion \cite{crocker_methods_1996,Shen2017,Novotny2019}. Figure \ref{Janus_traces} displays representative trajectories of Janus nanoparticles acquired at increasing levels of optical excitation power.
%The recorded two-dimensional (2D) particle trajectories are projections of the inherent three-dimensional (3D) motion onto the thin focal plane of the microscope objective, which has a finite depth of field. 
%Representative traces of Janus nanoparticles under increasing optical excitation power are shown in Figure \ref{Janus_traces}. 
The rapid rotational diffusion, characterized by the rotational diffusion coefficient  D$_{\rm{R}}$, reflects frequent reorientation events contributing to the stochastic nature of the trajectories (Fig.~\ref{Janus_traces}). For a quantitative analysis of the particle dynamics, we compute the ensemble-averaged Mean Squared Displacement (MSD) from approximately 100 individual traces acquired for each experimental condition (see \nameref{mat_methods} and Figs.~S1 and S2). The resulting MSDs are displayed in Fig.~\ref{MSD_linlin_AuNPs_Janus}.

The diffusion of a nanoparticle whose internal temperature exceeds that of the surrounding solvent is referred to as Hot Brownian Motion (HBM) \cite{Rings2010,Kroy2016,Sipova2020,Guerra2025}. The heat generated by the particle generates a radially symmetric thermal ``halo'' in the surrounding solvent, altering the local temperature and viscosity profiles around the particle and resulting in an apparent increase in both the rotational and translational diffusion coefficients. Despite its inherently non-equilibrium feature, %of thermally induced motion, 
the system can be described in a quasi-steady state (enabled by the separation of timescales between heat diffusion and particle motion) using equilibrium-like Stokes-Einstein relations with two distinct apparent temperatures and viscosities associated with the translational and rotational degrees of freedom, respectively. 
\cite{Rings2012,rings_theory_2011}. %Both translational and rotational diffusion coefficient are expected to increase with an increase of their internal temperature. *
For translational motion, the effective diffusion coefficient for a particle undergoing hot Brownian motion can then be written as:  %of a spherical particle of radius $R$ is: %with $k_\mathrm{B}$ the Boltzmann constant:  
\begin{equation}
\mathrm{D_{eff}^{HBM}}=\frac{k_B T_{\mathrm{eff}}}{6 \pi \eta_{\mathrm{eff}} R}
\label{D_HBM}    
\end{equation}
\noindent where the effective temperature is given by $T_{\mathrm{eff}}\simeq T_0(1+\frac{5}{12}\Delta T)$ with $\Delta T$ the temperature difference between the particle surface and the bulk solvent at temperature $T_0$, and the analytical expression of the effective viscosity $\eta_{\mathrm{eff}}$ is provided in the SI (Eq.~S4) \cite{Rings2010,rings_theory_2011,Kroy2016}.
The corresponding effective translational diffusion coefficient, $\mathrm{D_{eff}^{HBM}}$, which depends on the laser-induced temperature increase $\Delta T$, is related to the mean squared displacement in $n$ dimensions and at a given lag time $\Delta t$ by:
\begin{equation}
\label{MSD_long_time}
\mathrm{MSD}(\Delta t) = 2n \mathrm{D_{eff}^{HBM}} \cdot \Delta t,
\end{equation}
As we analyze the two-dimensional (2D) projection of the trajectories, we set $n=2$. Using the expressions of the effective temperature and viscosity (see SI), %we find that 
the rescaled diffusion coefficient can be explicitly calculated (Eq.~S5)\cite{rings_theory_2011} %(D$_{\mathrm{eff,Brownian}}$ - D$_0$)/D$_0$, 
and scales for small temperature increment, %to first order, 
linearly with $\Delta T$ (Fig.~S3), as:
\begin{equation}
\mathrm{\frac{D_{eff}^{HBM}-D_0}{D_0} }\propto \frac{\Delta T }{T_0},
\label{DeffBrownian}
\end{equation}
where D$_0$ is the diffusion coefficient in the absence of laser illumination, $\mathrm{D_0}=k_B T_0 / 6 \pi \eta(T_0) R$.
The temperature increase of an isolated %single 
plasmonic nanoparticle with absorption cross-section $\sigma_{abs} $ is given by:\cite{Baffou2010,Baffou2020}
\begin{equation}
\Delta T = \frac{\sigma_{abs} \mathrm{P}} {2 \pi^2 \kappa w^2 R}
    \label{TvsI}
\end{equation}
\noindent where P is the incident light power, $\kappa$ the thermal conductivity of the surrounding solvent ($\kappa$=0.6 W$\cdot$ m$^{-1}\cdot$K$^{-1}$ for water), and $w$ the beam waist. For an excitation wavelength of $\lambda =532$~nm, Mie theory yields an estimation of the absorption  cross-section of $\sigma_{abs} \approx 5000$~nm$^2$ for the Au nanoparticles used in this study. Combining Eqs.~\ref{DeffBrownian} and \ref{TvsI}, the contribution arising from hot Brownian motion is thus expected to have a linear dependence with the light intensity for a set of thermally independent particles: %\textit{i.e.} (D$\mathrm{_{eff}^{HBM}}$ - D$_0$)/D$_0 \propto I$.
\begin{equation}
\mathrm{\frac{D_{eff}^{HBM}-D_0}{D_0} }\propto \mathrm{P}.
\label{DeffHBM}
\end{equation}

For self-propelling active particles, such as the self-thermophoretic Janus nanoparticles studied here, the MSD acquires an additional contribution to account for the persistent and directed motion, as follows \cite{howse_self-motile_2007}:
\begin{equation}
\mathrm{MSD}(\Delta t) = 4 \mathrm{D_{eff}^{HBM}} \cdot \Delta t + \frac{v^2 \tau_\mathrm{R}^2}{2} \left [ \frac{2\Delta t}{\tau_\mathrm{R}} + e^{-2\Delta t/\tau_\mathrm{R}}-1 \right ],
\label{MSD_theo_expression_janus}
\end{equation}
where $v$ is the %constant 
propulsion speed, %$\mathrm{D_T}$ the translational diffusion coefficient due to thermal fluctuations, 
and $\tau_\mathrm{R}$ is the characteristic rotational diffusion time, inversely related to the rotational diffusion coefficient $\mathrm{D_R}$ by $\tau_\mathrm{R}$ = D$_\mathrm{R}^{-1}$. %= $8 \pi \eta R^3/k_\mathrm{B}T$. %with $R$ the radius of the particle, $k_\mathrm{B}$ the Boltzmann constant, and $T$ the absolute temperature.

The MSD behavior depends strongly on the timescale over which particle motion is probed. At long timescales, specifically when $\Delta t \gg \tau_\mathrm{R}$, as for our experimental conditions (see below), the MSD expression in Eq.~\ref{MSD_theo_expression_janus} reduces to a linear dependence in time, indicating an effective active diffusive regime: 
\begin{equation}
\label{longTime}
\mathrm{MSD}(\Delta t) \approx [4\cdot \mathrm{D_{eff}^{HBM}} + v^2 \tau_\mathrm{R}]\Delta t.
\end{equation}

This apparent diffusive behavior for active particles arises because rotational diffusion randomizes the propulsion direction over time, transforming persistent motion into a random walk leading to a substantial
enhancement of the effective diffusion coefficient over its
(hot) Brownian value, namely \cite{howse_self-motile_2007}:
\begin{equation}
\label{MSDactive}
 \mathrm{D_{eff}^{active} = D_{eff}^{HBM}} + \frac{1}{4} v^2 \tau_\mathrm{R}.
 \end{equation}
%In that case, the expression of the MSD in Equation \ref{MSD_theo_expression_janus} can be truncated to: $\mathrm{MSD}(\Delta t) = [4\cdot D_T + v^2 \tau_R]\Delta t$, and rewritten as:
%\begin{equation}
%\label{MSD}
%\mathrm{MSD}(\Delta t) = 4 D_{\mathrm{eff,active}} \cdot \Delta t ,
%\end{equation}
%where $ D_{\mathrm{eff,active}} = D_T + \frac{1}{4} v^2 \tau_R $.\\% For passive particles, there is no velocity and the usual expression of the MSD would be found.\\
This expression captures the enhanced diffusive behavior of active particles at long times combining effects of thermal fluctuations and self-propulsion.
On timescales longer than the characteristic rotational diffusion time, the motion resembles that of passive particles (Fig.~\ref{Janus_traces}), but with a markedly larger effective diffusion coefficient (Eq.~\ref{MSDactive}). 

This long timescale regime applies to our Janus nanoparticles, as the time range accessible in our experiments -- determined by the frame rate during video acquisition -- is 2.1~ms and above (see \nameref{mat_methods}). This is nearly one order of magnitude larger than the characteristic rotational time $\tau_\mathrm{R} \approx 0.22$~ms %(upper limit 
(calculated for a spherical particle with a diameter of 66~nm), confirming that $\Delta t \gg \tau_\mathrm{R}$.\\

Thus, for both Janus and bare gold nanoparticles, the MSD is expected to follow a diffusive regime (MSD $\propto \Delta t$) with an effective diffusion coefficient, D$_{\mathrm{eff}}$, differing only by the additional velocity contribution stemming from self-propulsion. 
Under illumination, the particle speed is expected to scale linearly with the laser power P and hence with the surface temperature increase $\Delta T$ (see Eq.~\ref{TvsI}),\cite{jiang_active_2010} leading, according to Eq.~\ref{MSDactive}, to a quadratic dependence of the normalized effective diffusion coefficient, 
\begin{equation}
\mathrm{\frac{D_{eff}^{active}-D_0}{D_0} }\propto a \cdot \mathrm{P} + b \cdot \mathrm{P}^2
%a\frac{\Delta T }{T_0} + b \left( \frac{\Delta T }{T_0} \right) ^2,
\label{DvsI}
\end{equation}
where $a$ and $b$ are two constants associated with hot Brownian motion and self-propulsion, respectively.

\begin{figure}
\centering
\includegraphics[width=\textwidth]{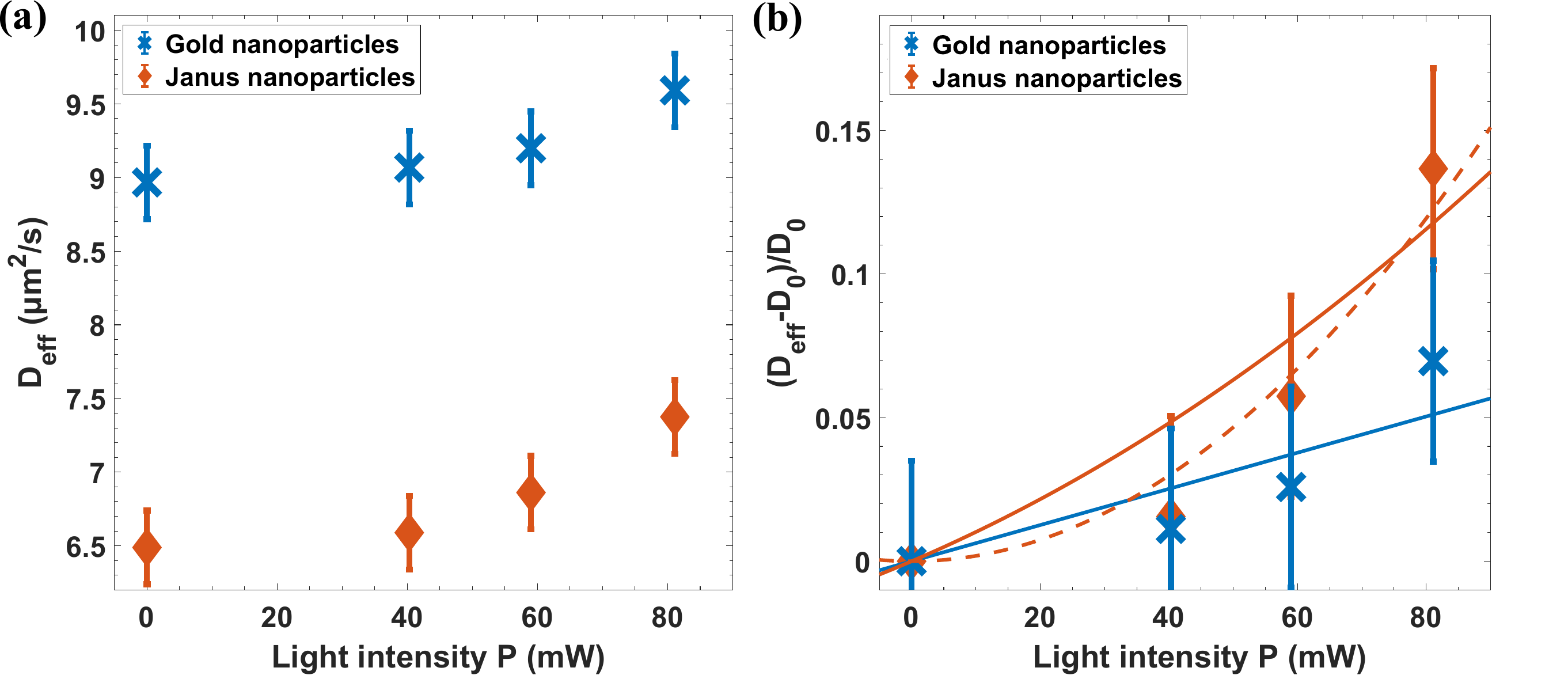}
\caption{\textbf{Experimental evidence of self-thermophoretic motion for Janus nanoparticles under optical illumination.} (a) Effective diffusive coefficients extracted from MSD analysis increase with light intensity. The lower absolute diffusion values of Janus nanoparticles, compared to bare Au particles, stem from their larger size due to the silica shell. (b) Normalized effective diffusion coefficients as a function of laser intensity. D$_0$ is the diffusion coefficient in absence of illumination. The enhanced dynamics of Janus nanoparticles reveal light-activated self-propulsion in addition to hot Brownian motion, which is also observed for Au nanoparticles (linear fit (Eq.~\ref{DeffHBM}) shown as a blue line). Orange lines represent quadratic fits for Janus nanoparticles: the dashed curve is purely quadratic (Eq.~\ref{DvsI}, $a=0$), while the solid curve includes the same linear contribution as for Au nanobeads (Eq.~\ref{DvsI}, $a=6.35\times 10^{-4}$, $b=9.85\times 10^{-6}$). Error bars represent estimated experimental uncertainties.  
}
\label{Deff_raw_rescaled}

\end{figure}

The experimental averaged MSDs for both Janus and bare gold nanoparticles are shown in Figure~\ref{MSD_linlin_AuNPs_Janus}, and linear fits of these curves provide the  corresponding effective diffusion coefficients, which are presented in Figure~\ref{Deff_raw_rescaled}(a).
For the samples without optical excitation, the fits yield diffusion coefficients of D$_{0,Au} =8.98 \pm0.25$~µm$^2$/s and D$_{0,Janus} = 6.49 \pm 0.25$~µm$^2$/s (Figs.~S1 and S2). Using the Stokes-Einstein equation, these values correspond %at \SI{25}{\degreeCelsius}, 
to hydrodynamic diameters of $d_{H,Au} = 54.6 \pm $~1.5~nm and $d_{H,Janus} = 75.6 \pm$ ~2.9~nm, respectively. As expected, these hydrodynamic diameters are larger than the bares core sizes $d$ obtained from the size distribution measured by TEM (Fig.~\ref{size_distribution}), and are consistent with $d_{H} \sim d + \kappa^{-1} $, where $\kappa^{-1}$=13~nm is the Debye screening length at the experimental ionic strength (I=0.5~mM) used in our nanoparticle suspensions (see \nameref{mat_methods}).
%due to particle-solvent interactions that lead to a reduced diffusion and consequently an increased effective size.\\

Characterizing the active behavior of Janus nanoparticles under illumination solely from their MSD is challenging, as it requires precise quantification of the light-induced enhancement in both translational and rotational diffusion (Eqs.~\ref{MSD_theo_expression_janus} and \ref{MSDactive}) \cite{xuan_near_2016,Bailey2022}. To isolate the contribution of active motion and disentangle it from hot Brownian motion, we compare the dynamics of the Janus nanoparticles with that of bare gold nanoparticles under identical experimental conditions and same laser illumination. 
To account for the size difference caused by the silica shell on the Janus particles, we analyze the rescaled effective translational diffusion coefficient, (D$_{\mathrm{eff}}$ - D$_0$)/D$_0$, where D$_0$ is the diffusion coefficient in the absence of laser illumination. This parameter quantifies the relative increase in the effective diffusion coefficient for both systems, as shown in Figure \ref{Deff_raw_rescaled}(b). 
If the observed enhancement was solely due to hot Brownian motion, the increase in %translational and rotational 
diffusion due to internal heating of the Janus particles should not exceed that of the bare gold nanoparticles. However, our results reveal that the relative increase of the normalized effective diffusion coefficient is consistently higher for the Janus nanoparticles than for the bare gold nanoparticles, demonstrating that Janus particles exhibit significant self-propulsion. 

The quantitative analysis is provided below. The dependence of Au nanoparticles on light intensity deviates from linearity at the highest laser power, suggesting that collective thermal effects contribute to their heating.  
When several nanoparticles are illuminated at the same time,  the temperature increase experienced by a nanoparticle also stems from neighboring nanoparticles heating their environment.\cite{Baffou2020} This effect originates from the long-range temperature diffusion
profile around a source of heat, decaying as $\Delta T(r) \propto 1/r, r$ being
the distance from the heat source.\cite{Rings2010} %Eq.~\ref{TvsI} leads to a substantial underestimation of actual heating in the sample as revealed by the non-linearity of the HBM diffusion increase for Au nanoparticles (Fig.~\ref{Deff_raw_rescaled}).
As shown in Fig.~S4, the temperature increase calculated from the experimental rescaled diffusion coefficient (Fig.~\ref{Deff_raw_rescaled}(b)) using the HBM theory (Eq.~S5 and Fig.~S3) consistently exceeds the prediction of Eq.~\ref{TvsI} for thermally independent particles. This discrepancy persists even when considering the lower limit of the beam size, which corresponds to the upper limit of the laser power density (Fig.~S4).

Conversely, for Janus nanoparticles, a non-linear dependence of the normalized effective diffusion coefficient on laser power is expected from Eq.~\ref{DvsI}, in agreement with the experimental results shown in Figure \ref{Deff_raw_rescaled}(b). 
To isolate the contribution of active motion, we have used in Eq.~\ref{DvsI} the same fitting parameter accounting for HBM as that determined for bare gold nanoparticles (namely parameter $a$), while allowing the active term $b$ to vary freely.
Although the fit quality remains limited, the pronounced increase in the normalized effective diffusion coefficient provides conclusive evidence of an additional contribution from self-thermophoresis–induced motion to the dynamics of Janus nanoparticles. At the highest illumination, active motion accounts for approximately 50\% of the total diffusion enhancement.
The corresponding propulsion speed of the Janus particles can then be estimated using Eq.~\ref{MSDactive}, assuming $\tau_\mathrm{R}$ = D$_\mathrm{R}^{-1}(T_{\mathrm{eff}}) \approx $D$_\mathrm{R}^{-1}(T_0)$, which yields $v \sim 35$~µm/s. The corresponding P\'eclet number 
can also be calculated and gives $Pe \approx 0.3$, confirming \textit{a posteriori} that the system operates in a regime where active and passive Brownian contributions are comparable.

\clearpage

\section{Conclusion}

In summary, we have successfully produced stable nanoscale gold-silica Janus particles and demonstrated their self-thermophoretic propulsion under optical illumination. Using single particle tracking, we resolved individual particle trajectories, and determined effective diffusion coefficients by analyzing the mean squared displacements of approximately one hundred individual traces per experimental condition, ensuring statistical relevance. %\cite{Novotny2019}.
The limited amplitude of the active contribution within the explored irradiation power range, combined with the intrinsic passive hot Brownian diffusion of the nanoparticles, results in an experimental system at low P\'eclet number in which active and passive contributions to particle dynamics are comparable in magnitude. In this low P\'eclet regime, our results reveal that the Janus particles exhibit a systematically and significantly higher relative increase in normalized effective diffusivity compared to bare gold nanobeads under identical illumination conditions. This enhancement, which accounts up to about 50\% of the total dynamics, provides direct experimental evidence of visible-light-induced active motion in Janus nanoparticles.
Furthermore, since the active behavior is both optically controlled and reversible, our system represents a simple, fuel-free, tunable platform for studying active matter and non-equilibrium dynamics at the nanoscale. 

\clearpage

%%%%%%%%%%%%%%%%%%%%%%%%%%%%%%%%%%%%%%%%%%%%%%%%%%%%%%%%%%%%%%%%%%%%%
%% The "Acknowledgement" section can be given in all manuscript
%% classes.  This should be given within the "acknowledgement"
%% environment, which will make the correct section or running title.
%%%%%%%%%%%%%%%%%%%%%%%%%%%%%%%%%%%%%%%%%%%%%%%%%%%%%%%%%%%%%%%%%%%%%

\section{Author contributions}
H.T. performed the experiments, acquired and analyzed the data, and wrote the first version of the manuscript. C.M. and B.A. synthesized the nanoparticles. L.B. and H.T. developed and assembled the optical setup. E.G. conceived and supervised the project, analyzed and interpreted the data, secured funding, wrote and edited the manuscript. All authors reviewed and approved the final version of the manuscript.

\section{Conflicts of interest}
There are no conflicts to declare.

\begin{acknowledgement}
We thank Alois W\"urger for useful discussion. We acknowledge financial support from the French National
Research Agency (ANR) under Grants No. ANR-19-CE30-0024 ``ViroLego" and ANR-18-CE09-0025 ``SoftQC".
\end{acknowledgement}

%%%%%%%%%%%%%%%%%%%%%%%%%%%%%%%%%%%%%%%%%%%%%%%%%%%%%%%%%%%%%%%%%%%%%
%% The same is true for Supporting Information, which should use the
%% suppinfo environment.
%%%%%%%%%%%%%%%%%%%%%%%%%%%%%%%%%%%%%%%%%%%%%%%%%%%%%%%%%%%%%%%%%%%%%
\section{Supplementary Information (SI) available}
The Supplementary Information includes 
%absorption spectra of the synthesized nanoparticles and their size distribution determined by TEM; a schematic representation of the optical setup; a representative dark-field microscopy image illustrating particle scattering; validation of uniform illumination confirming the absence of gradient-induced artifacts; and a complete 
the detailed dynamics analysis comprising mean squared displacement (MSD) curves for all individual trajectories of both Au and Au-SiO$_2$ nanoparticle systems under each experimental condition, the theoretical expressions and plot of the normalized diffusion coefficient and viscosity for hot Brownian motion, as well as the corresponding temperature increase as a function of laser illumination. A dark-field microscopy movie of Janus nanopaticles in absence of laser illumination is provided (AVI).  

%The characterization of the particles is provided through their absorption spectrum and their size distribution determined by TEM in the Supporting Information. Also included are the details on the experimental optical setup. An illustration of a typical image obtained via Dark-Field microscopy is provided, as well as the verification of the homogeneous illumination and the absence of any gradient-related effects, and the MSD bundles for each experimental condition for both nanoparticle (Au and Au-SiO$_2$) systems. 

%%%%%%%%%%%%%%%%%%%%%%%%%%%%%%%%%%%%%%%%%%%%%%%%%%%%%%%%%%%%%%%%%%%%%
%% The appropriate \bibliography command should be placed here.
%% Notice that the class file automatically sets \bibliographystyle
%% and also names the section correctly.
%%%%%%%%%%%%%%%%%%%%%%%%%%%%%%%%%%%%%%%%%%%%%%%%%%%%%%%%%%%%%%%%%%%%%

\clearpage
\bibliography{Janus_NPs_article}

\end{document}